\documentclass[11pt]{article}
\usepackage{amsfonts}
\usepackage{amssymb}
\usepackage{amsmath}
\newcommand{\nc}{\newcommand}

\nc{\beq}{\begin{equation}}
\nc{\eeq}{\end{equation}}
\nc{\barray}{\begin{eqnarray}}
\nc{\earray}{\end{eqnarray}}
\nc{\barrayn}{\begin{eqnarray*}}
\nc{\earrayn}{\end{eqnarray*}}
\nc{\bcenter}{\begin{center}}
\nc{\ecenter}{\end{center}}
\nc{\ket}[1]{| #1 \rangle}
\nc{\bra}[1]{\langle #1 |}
\nc{\0}{\ket{0}}
\nc{\mc}{\mathcal}
\nc{\er}[1]{(\ref{eq:#1})}
\nc{\onehalf}{\tfrac{1}{2}}
\nc{\partialbar}{\bar{\partial}}
\nc{\Tr}{\mbox{Tr}}
\nc{\diagonal}[1]{\mbox{Diag}\left(#1\right)}
\nc{\tl}{\tilde{\lambda}}
\nc{\tH}{\tilde{H}}
\nc{\psit}{\widetilde{\psi}}
\nc{\chit}{\widetilde{\chi}}
\nc{\tO}{\widetilde{\mc{O}}}
\nc{\mcv}{\mc{V}}
\nc{\mcc}{\mc{C}}
\nc{\mco}{\mc{O}}
\nc{\mca}{\mc{A}}
\nc{\tnu}{\tilde{\nu}}
\nc{\tmu}{\tilde{\mu}}
\nc{\tN}{\tilde{N}}
\nc{\tM}{\tilde{M}}
\nc{\dda}{{\partial\over\partial a}}
\nc{\infinity}{\infty}
\nc{\sign}[1]{\mbox{sign}\left( #1 \right)}
\nc{\openone}{{\mathbb 1}}
\nc{\openz}{{\mathbb Z}}

\renewcommand{\thefootnote}{\fnsymbol{footnote}}
\begin{document}
\begin{titlepage}
{\flushright{\small MIT-CTP-3556 \\hep-th/0411040\\}}

\begin{center}
{\LARGE\bf Closed Superstring Emission from Rolling Tachyon Backgrounds}

\vspace{5mm}
{\sc Jessie Shelton}\footnote{E-mail: jshelton@mit.edu.  }

{\it Center for Theoretical Physics\\  Massachusetts 
Institute of Technology\\ Cambridge, MA 02139, U.S.A.}

\end{center}

\vspace{1cm}
\abstract{We compute the lowest components of the Type II Ramond-Ramond boundary state for the tachyon profile $T (X) = \lambda e ^{X ^ 0/\sqrt{ 2 }}$ by direct path integral evaluation.  The calculation is made possible by noting that the integrals involved in the requisite disk one-point functions reduce to integrals over the product group manifold $U (n)\times U (m)$.  We further note that one-point functions of more general closed string operators in this background can also be related to  $U (n)\times U (m)$ group integrals.  Using this boundary state, we compute the closed string emission from a decaying unstable D$p$-brane of Type II string theory.  We also discuss closed string emission from the tachyon profile $T (X) =\lambda\cosh (X ^ 0/\sqrt{ 2 })$. We find in both cases that the total number of particles produced diverges for  $p = 0$, while the energy radiated into closed string modes diverges for $p\leq 2$, in precise analogy to the bosonic case.}

\end{titlepage}

\renewcommand{\thefootnote}{\arabic{footnote}}
\setcounter{footnote}{0}

%%%%%%%%%%%%%%%%%%%%%%%%%%%%%%%%%%%%%%%%%%%%%%%%%%%%%%%%%%%%%%%%%%%%%%%%%%%%%%%%%%%%%
\section{Introduction}
%%%%%%%%%%%%%%%%%%%%%%%%%%%%%%%%%%%%%%%%%%%%%%%%%%%%%%%%%%%%%%%%%%%%%%%%%%%%%%%%%%%%%

The worldsheet approach to studying tachyon condensation was introduced by Sen in \cite{sen-rt, sen-tm}.  In this approach, we give the tachyon a space-time-dependent expectation value by deforming the worldsheet conformal field theory with the exactly marginal boundary operator $T (X) =\lambda\cosh X ^ 0$, or, in the supersymmetric theory, the operator ${\lambda\over\sqrt{ 2 }}\eta\psi ^ 0 \sinh (X ^ 0/\sqrt{ 2 })$, which corresponds to the tachyon profile $T (X) =\lambda\cosh (X ^ 0/\sqrt{ 2 })$.  Another tachyon profile $T (X) = \lambda e ^{X ^ 0}$ and its supersymmetric analogue $T (X) = \lambda e ^{X ^ 0/\sqrt{ 2 }}$ was introduced in \cite{lnt}.  These conformal field theories are related by Wick rotation to theories where the boundary deformation is taken from a local SU(2) current algebra. 

Worldsheet conformal field theory is a perturbative approach to string theory and can likely provide only a partial answer to the deep nonperturbative problems posed by D-brane decay.  Working out conformal field theory descriptions of D-brane decay is nonetheless of interest.  Perhaps most importantly, the conformal field theory approach provides a clear computation of the coupling of the decaying D-brane to closed strings, a problem which remains somewhat obscure in the open string field theory description of tachyon condensation.  In the worldsheet approach, the final closed string state produced by the decaying brane can be readily computed \cite{llm}.

The rolling tachyon background is also a good starting point for exploring some of the technical issues involved in working with time-dependent perturbative solutions to the string equations of motion, as the time dependence is confined to the open string sector, and thereby poses fewer new conceptual challenges than working with a time-dependent space-time background.  One such interesting technical issue is the relationship proposed in \cite{gs} between the mathematical ambiguities involved in defining the correlators in a nontrivial wrong sign CFT and the physical ambiguities involving choice of vacua in time-dependent field theories.  

Although it appears that the perturbative worldsheet description of D-brane decay breaks down in most examples, as the number of emitted particles and amount of emitted energy becomes infinite \cite{llm}, there are examples where this divergence is brought under control, either due to a higher Hagedorn temperature \cite{klms}, or an ability to work with a dual, non-perturbative description of the system \cite{kms}. 

In Type II theories, the space-time stress energy tensor for a decaying brane has been studied in \cite{sen-tm, lnt}.  In this note we present a discussion of closed string production from a decaying Type II D$p$-brane.

The marginal deformations we study are the so-called ``rolling tachyon''
\footnote{There has been some confusion in the literature as to the possible appearance of a term of the form $T ^ 2$ in the boundary action.  If the boundary action is written in the manifestly supersymmetric form $S_{bdy} =\int dt\, d\theta (\Gamma D\Gamma +\Gamma T (Y))$ where $\Gamma =\eta +\theta F$, $D =\partial_{\theta} +\theta\partial_t$, and $Y = X+\theta\psi$, a term of the form $T ^ 2$ arises after elimination of the auxiliary field $F$.  This form of the boundary action has been instrumental in studies of tachyon condensation using boundary string field theory.  However, when the boundary theory is defined through deformation by the operators above, the $T ^ 2$ term does not arise; its contribution to the boundary action vanishes on shell.  It is simple to check that the supersymmetry Ward identities for the deformed SCFT are satisfied when the boundary action only contains the term $\eta\psi T' (x)$.}  
\beq
\label{eq:rolling-tachyon}
\Delta S =-\sqrt{ 2 }\pi\lambda\int {dt\over 2\pi}\eta\psi ^ 0\, e ^{X ^ 0/\sqrt{ 2 }}
\eeq
and the ``bouncing tachyon''
\beq
\label{eq:bouncing-tachyon}
\Delta S =-\sqrt{ 2 }\pi\lambda\int {dt\over 2\pi}\eta\psi ^ 0\,\sinh\left({X ^ 0\over\sqrt{ 2 }}\right).
\eeq
Here $\eta$ is a fermionic degree of freedom living on the boundary of the worldsheet; it has trivial worldsheet dynamics, and may also be understood as a Chan-Paton factor.  It appears in the vertex operators of all open string states with wrong sign GSO projection.  The coupling $\lambda$ is related to the initial position and velocity of the tachyon at the top of its potential \cite{sen-rt}.

For most of our calculations, we will assume that we are able to choose a gauge in which no timelike oscillators are excited.  Then a general closed string vertex operator takes the form
\beq
\label{eq:vns}
\mcv_s = e ^{i E_s X ^ 0} \mcv ^ {\perp}_s (X ^ i,\psi ^ i,\psit ^ i)
\eeq
in the NS-NS sector and
\beq
\label{eq:vr}
\mcv_s =e ^{i E_s X ^ 0} \Theta_{s_0} \widetilde{\Theta}_{s'_0}\mcv ^ {\perp}_s (X ^ i,\psi ^ i,\psit ^ i)
\eeq
in the R-R sector.  Here the $\Theta_{s}$ are spin fields, $\Theta_{s_0} = e ^{is_0H_0}$. With this gauge fixing, the problem of computing general closed string one point functions in the rolling tachyon background is greatly simplified.  We only need to know the expectation values
\[
\langle e ^{-\Delta S}\, e ^{i E_s X ^ 0}\rangle_{disk}
\]
in the NS-NS sector, and
\[
\langle e ^{-\Delta S}\,e ^{i E_s X ^ 0} \Theta_{s_0} \widetilde{\Theta}_{s'_0}\rangle_{disk}
\]
in the R-R sector.  Equivalently, we only need to know the lowest component of the boundary state describing the deformed $X ^ 0$ CFT,
\[
\ket{B} ^ 0 = g (x ^ 0)\0+\ldots.
\]
In the NS-NS sector, the function $g (x ^ 0)$ for the tachyon profile \er{bouncing-tachyon} was worked out in \cite{sen-tm} and for the tachyon profile \er{rolling-tachyon} in \cite{lnt}.

The computation of $g (x ^ 0)$ for both tachyon profiles \er{rolling-tachyon} and \er{bouncing-tachyon} in the R-R sector is one of the aims of this paper.\footnote{See also \cite{sen-guess}.}  We will, however, spend more time on the rolling tachyon profile, \er{rolling-tachyon} above, as a direct perturbative calculation of the closed string disk one-point amplitudes in that background reveals an interesting $U (p)\times U (q)$ structure, which allow the correlators to be easily calculated.  This structure is an interesting generalization of the $U (n)$ symmetry that was found in the bosonic case in \cite{lnt, Neil}.  We will also compute the NS-NS one-point functions for this background in order to demonstrate a similar $U (p)\times U (p)$ structure appearing in the correlators there.  These calculations are in section 2.

In section 3, we discuss how the relation between correlation functions in the rolling tachyon background and integrals of invariants over the group $U (p)\times U (q)$ may be extended to one-point functions of more general operators.

In order to compute the function $g (x ^ 0)$ for the bouncing tachyon, we must use some more sophisticated CFT machinery, namely the SU(2) current algebra of the Euclidean version of the theory. This technique is the one originally used by Sen. We have placed this computation in appendix A and use only the result, equation \er{fdef}, in the main body of the paper.  Also in this appendix, we present an alternate derivation of the function $g (x ^ 0)$ for the rolling tachyon. 

Next, we consider the closed string production from the decaying D-brane, for both tachyon profiles.  As in the bosonic case \cite{llm}, both the number and the energy of emitted particles diverge for low dimensional D-branes.  Since the tachyon profiles \er{rolling-tachyon} and \er{bouncing-tachyon} describe homogenous tachyon condensation in the $p$ spatial dimensions of the brane, the condensation process is unphysical if $p\neq 0$.  A more physical decay process for a higher-dimensional D-brane will likely proceed in several causally disconnected patches, each of which behaves like a decaying D-brane of lower dimension \cite{lnt}.  Thus we expect that in a physical decay process, the number and energy of particles emitted from decaying D$p$-branes will diverge for any $p$.  This suggests that all the energy of the D-brane is converted into closed strings.

Finally, appendix B summarizes the relationship between two different bases for the spinor vacua, the states $\ket{s}\ket{\tilde{s}}$ in terms of which the R-R spin fields have a simple description, and the states $\ket{\pm}$ in terms of which the boundary state has a simple description.  These relationships will be useful when we compute the R-R sector boundary state.

%%%%%%%%%%%%%%%%%%%%%%%%%%%%%%%%%%%%%%%%%%%%%%%%%%%%%%%%%%%%%%%%%%%%%%%%%%%%%%%%%%%%%
\section{Disk One-Point Functions in the Rolling Tachyon Background}
%%%%%%%%%%%%%%%%%%%%%%%%%%%%%%%%%%%%%%%%%%%%%%%%%%%%%%%%%%%%%%%%%%%%%%%%%%%%%%%%%%%%%

In this section, we compute disk one-point functions of the form
\[
\left\langle\Theta_s\widetilde{\Theta}_{s'}\right\rangle_{deformed} =
\left\langle e ^{-\Delta S}
\Theta_s\widetilde{\Theta}_{s'}\right\rangle_{free}
\]
by Taylor-expanding the exponential and evaluating the correlators that appear at each order in $\lambda$. We do not carry out the path integral over the $X ^ 0$ zero mode $x ^ 0$ in the expectation values here, but interpret the resulting amplitudes as a function of $x ^ 0$.  We therefore define $\tl\equiv\lambda e ^{x ^ 0/\sqrt{2}}$, and
\barray
\nonumber
\left\langle\exp \left( \sqrt{ 2 }\pi\lambda\int {dt\over 2\pi}\, \eta\psi ^ 0 e ^{X ^ 0/\sqrt{ 2 }}\right)
\Theta_s\widetilde{\Theta}_{s'}\right\rangle
& = &\sum_n{(\sqrt{ 2 }\pi\tl) ^ n\over n! }
       \left\langle \left(\int {dt\over 2\pi}\, \eta\psi ^ 0 e ^{X ^ 0/\sqrt{ 2 }}\right) ^ n
\Theta_s\widetilde{\Theta}_{s'}\right\rangle \\
\label{eq:adef}
&\equiv &\sum_n (i\pi\tl) ^ n\mca_n . 
\earray
In the R-R sector, the boundary fermion $\eta$ has a zero mode; thus, only odd numbers of insertions of the boundary perturbation give non-vanishing contributions.  Since the worldsheet dynamics of the boundary fermion are trivial, it is simply a matter of convenience whether we choose to formally integrate it out, as in \cite{lnt}, or include it in the vertex operator, as in \cite{sen-tm}.  We will find it convenient to include $\eta$ in the vertex operator, since in this case the vertex operators on the boundary all commute with each other.

The structure of this computation is most evident when using the bosonized representation of the fermion fields.  The correlators that appear in  \er{adef} are then
\[
\frac{ (\pi\tl) ^{2k+1} }{ (2k+1)!} \int\prod_{i = 1} ^{2k+1}\frac{dt_i}{2\pi}
\left\langle \left[e^{iH_0 }(t_i)-e^{-iH_0 } (t_i)\right]\, e^{X_0 /\sqrt{2}} (t_i) \,
e ^{is\cdot H} (0)  e ^{is'\cdot \tH} (0 )\right\rangle  .
\]
This expectation value factors into three parts.  The contribution from the transverse fermions is $\delta_{\vec{s}, - \vec{s}'}$, which we will drop for ease of notation.  The bosonic contribution is simply
\beq
\label{eq:boson-correlators}
\left\langle \prod_{i = 1} ^{2k+1} e^{X_0 /\sqrt{2}}(t_i) \right\rangle =
\prod_{i < j} | e ^{it_i} -e ^{it_j} | =\prod_{i < j} 2\sin\left({t_i-t_j\over 2}\right).
\eeq
Now consider $\left\langle\prod_{i = 1} ^{2k+1}\left[e^{iH_0 }(t_i)-e^{-iH_0 }(t_i) \right]
e ^{is_0 H_0} (0) e ^{is'_0\tH_0} (0 )\right\rangle$.  From now on we will drop the label 0 to clean up the notation.  When we expand the product $\prod_{i = 1} ^{n}(e^{iH }(t_i)-e^{-iH }(t_i) )$, we will find a sum of terms, each of which has the form $\prod_{i = 1} ^{2k+1} e ^{i\epsilon_i H} (t_i)$, where $\epsilon_i =\pm 1$. The path integral over the $H$ zero mode imposes $H$ momentum conservation, giving the condition
\[
s+s' +\sum_{i = 1} ^ n\epsilon_i = 0,
\]
where $\epsilon_i =\pm 1$.  Since $s, s' =\pm\onehalf$, we have two possible ways to satisfy the condition above: let $s=s'=\pm\onehalf$ and retain only terms in the expansion of the product $\prod_{i = 1} ^{2k+1}(e^{iH }(t_i)-e^{-iH }(t_i) )$ where the $H$ momentum sums to $\mp 1$.  
There will be a sign difference between the expectation values in the two different cases, as when we choose $s = s' =\onehalf $, we pick up an odd number of factors of $- e ^{-i H}$, and an even number when we choose $s = s' =-\onehalf $.  As we will see, this sign difference gives us the correct GSO projection.

Let us consider the $s = s' =\onehalf$ for definiteness.  Then in the product $\prod_{i = 1} ^{2k+1}\left[e^{iH }(t_i)-e^{-iH }(t_i) \right]$ we need to keep the terms with $k+1 $ minus signs, that is, terms with $\sum\epsilon_i =-1$.  
The $H$ correlators become
\[
(i) ^{- k}
e^{- i\sum t_i/2} \prod_{i < j} 2\sin \left( \frac{t_i -t_j}{2}\right)^{\epsilon_i\epsilon_j}.
\]
Including the phases coming from the transformation from $z$ to $t$ thus yields simply
\beq
\label{eq:fermion-correlators}
(- 1) ^{k} \sqrt{i} \prod_{i < j}2\sin \left( \frac{t_i -t_j}{2}\right)^{\epsilon_i\epsilon_j}.
\eeq

Combining equations \er{boson-correlators} and \er{fermion-correlators}, we have (adjusting the overall phase for simplicity)
\beq
\label{eq:term-n}
\mca_{2k+1} =
\frac{1 }{ (2k+1)!} \int\prod_{i = 1} ^{2k+1}\frac{dt_i}{2\pi} \sum_{\sigma\in P}\left[
\prod_{\sigma (i) < \sigma (j)} \left( 2\sin \left( \frac{t_{\sigma ( i)}-t_{\sigma (j)} }{2} \right) \right) ^{1+\epsilon_{\sigma (i)}\epsilon_{\sigma (j)}} \right],
\eeq
where $P$ denotes the set of all distinct ways to distribute $k+1 $ values of $-1$ and $k $ values of $1$ among the $\epsilon_i$. 
Since the boundary operators are not path ordered, each of the $\left(\begin{array}{c} 2k+1\\k\end{array}\right)$ terms in the sum above contribute equally.  Thus, we may choose one ordering of the $\epsilon_i$, namely
\[
\epsilon_i =-1,  \phantom{mm} 1\leq i \leq k+1;\phantom{spacer spacer}
\epsilon_i = 1,\phantom{mm} k+1 < i\leq 2k+1,
\]
and write the correlator in terms of that ordering alone:
\beq
\label{eq:rhinoceros}
\mca_{2k+1} =
\frac{1 }{ \left(k+1\right)!k!}  \int\prod_{i = 1} ^{2k+1}\frac{dt_i}{2\pi} 
\prod_{i < j} \left( 2\sin \left( \frac{t_i-t_j}{2} \right) \right) ^{1+\epsilon_i\epsilon_j} .
\eeq
Now, the Haar measure for the group $U (n)$ is
\[
\int_{U (n)} [d U] \; {\bf 1} ={1\over n!} \int\prod_{i = 1} ^{n}\frac{dt_i}{2\pi} \Delta ^ 2 (t)
= 1,
\]
where $\Delta (t)$ is the Vandermonde determinant,
\[
\Delta  (t) =\prod_{i < j} ^ n2\sin \left({t_i-t_j\over 2}\right).
\]
It is now easy to see that \er{rhinoceros} is simply the integral of unity over the group manifold of the product group $U (k+1)\times U (k)$:
\[
\mca_{2k+1} =\int_{U (k+1)} [d U]\, {\bf 1}\;\int_{ U (k)} [d U]\, {\bf 1} = 1.
\]
Since $\mca_{2k+1} = 1$ for all $k $, the sum over all orders \er{adef} becomes simply 
\beq
\label{eq:gr-def}
\sum_{k} (i\pi\tl) ^{2k+1}
=\frac{i\pi\lambda \, e ^{x ^ 0/\sqrt{2}}}{1+\pi ^ 2\lambda ^ 2 e ^{\sqrt{2}x ^ 0}}\equiv g ^ R (x ^ 0).
\eeq
It remains to restore the dependence on the spin.

The amplitudes that we have calculated are built on spin vacua of the form $(\ket{\onehalf}\ket{\onehalf} -\ket{{- \onehalf}}\ket{{- \onehalf}} )$ where we have labeled states by their $s_0$ quantum numbers, suppressing the transverse spin eigenvalues.  The R-R boundary state, however, is built on spinors $\ket{\pm\pm}$ that satisfy $(\psi ^ 0_0\pm i\psit ^ 0_0)\ket{\pm\pm} =(\psi ^ 1_0\pm i\psit ^ 1_0)\ket{\pm\pm} = 0$.  Thus to write down the boundary state, we need to know the relationship between the two bases $\ket{s}\ket{s'}$ and $\ket{\pm,\pm}$, which we work out in appendix B. We find from \er{spin-timelike}
\[
\ket{\onehalf}\ket{\onehalf} -\ket{{- \onehalf}}\ket{{- \onehalf}} =\ket{+-} +\ket{-+}.
\]  
We may now easily restore the transverse spin quantum numbers to find that the spin vacuum should be $\ket{+-\cdots -} + \ket{- + \cdots +}$.  Note that this is the correct GSO projection for Type IIA, and has the right Lorentz properties to source $C_9$ flux. The boundary state for the supersymmetric rolling tachyon profile is therefore
\beq
\ket{B}_R = g ^ R (x ^ 0)\left(\ket{+-\cdots -} + \ket{- + \cdots +}\right) +\mbox{oscillators}.
\eeq
The generalization to other D$p$-branes of either Type II string theory is very simple; we simply need to change the dependence on the transverse spin degrees of freedom, exactly as we would in the absence of the tachyon expectation value.

%%%%%%%%%%%%%%%%%%%%%%%%%%%%%%%%%%%%%%%%%%%%%%%%%%%%%%%%%%%%%%%%%%%%%%%%%%%%%%%%%%%%%
\subsection{NS-NS sector}
%%%%%%%%%%%%%%%%%%%%%%%%%%%%%%%%%%%%%%%%%%%%%%%%%%%%%%%%%%%%%%%%%%%%%%%%%%%%%%%%%%%%%

In the NS-NS sector, the expectation value that we need to compute is
\[
\left\langle{\bf1}\right\rangle_{deformed} =\left\langle e ^{-\Delta S}\right\rangle_{free}
\]
This expectation value was discussed in \cite{lnt}; our principal interest in this section will be to demonstrate an interesting $U (n)\times U (n)$ structure that allows the integrals to be easily performed.  As before, we will evaluate this correlator by expanding in powers of the boundary perturbation, and define
\[
\left\langle \exp\left( \sqrt{ 2 }\pi\lambda\int {dt\over 2\pi}\, \eta\psi ^ 0 e ^{X ^ 0/\sqrt{ 2 }}\right)
\openone\right\rangle =\sum_n (i\pi\tl) ^ n\mca_n.
\]
The boundary fermion $\eta$ has no zero mode in the NS-NS sector, so only terms in the expansion with $n$ even will contribute.

Using bosonized fermions, $\mca_{2k} $ is
\beq
\label{eq:fish}
\mca_{2k} =
\frac{1 }{ (2k)!} \int\prod_{i = 1} ^{2k} \frac{dt_i}{2\pi}
\left\langle \left[e^{iH }(t_i)-e^{-iH } (t_i)\right]\, e^{X_0 /\sqrt{2}} (t_i) \,
\right\rangle  .
\eeq
The $H$-momentum conservation condition now yields 
\[
\sum_i\epsilon_i = 0.
\]
The expectation value \er{fish} is thus
\[
\mca_{2k} =
{1\over (2k)!} \int\prod_{i = 1} ^{2k}\frac{dt_i}{2\pi} \sum_{\sigma\in P}\left[
\prod_{\sigma (i) < \sigma (j)} \left( 2\sin \left( \frac{t_{\sigma (i)}-t_{\sigma (j)} }{2} \right) \right) ^{1+\epsilon_{\sigma (i)}\epsilon_{\sigma (j)}}\right],
\]
where $P$ denotes the set of all distinct ways to distribute $k$ values of 1 and $k$ values of $-1$ among the $\epsilon_i$.  Again we choose a representative ordering,
\[
\epsilon_i =-1,  \phantom{mm} 1\leq i\leq k;\phantom{space spacer r}
\epsilon_i = 1,\phantom{mm} k < i\leq 2k,
\]
and write the correlator in terms of that ordering alone:
\beq
\label{eq:moose}
\mca_n =
\frac{1 }{ (k!) ^ 2}  \int\prod_{i = 1} ^{2k}\frac{dt_i}{2\pi} 
\prod_{i < j} \left( 2\sin \left( \frac{t_i-t_j}{2} \right) \right) ^{1+\epsilon_i\epsilon_j} .
\eeq
This can easily be written as the integral of unity over the group manifold of the product group $U (k)\times U (k)$:
\[
\mca_{2k} =
\left[\int_{U (k)} [d U]\, {\bf 1}\right] ^ 2= 1
\]
for all $k $.
Summing all orders,  the functional form of the time-dependent NS-NS source is then
\beq
\label{eq:gns-def}
\sum_{k} (i\pi\tl) ^ {2k}
=\frac{1 }{1+\pi ^ 2\lambda ^ 2 e ^{\sqrt{2}x ^ 0}}\equiv g ^{NS} (x ^ 0).
\eeq
The boundary state in the NS-NS sector is therefore
\beq
\ket{B}_{NS} = g ^{NS} (x ^ 0)\left(\ket{0; +}-\ket{0;-}\right) + \mbox{oscillators}.
\eeq
This answer differs from that found in \cite{lnt} by $\lambda_{here} =\sqrt{2}\lambda_{there}$.  This difference is not meaningful as $\lambda$ can be set to any value by a time translation.

%%%%%%%%%%%%%%%%%%%%%%%%%%%%%%%%%%%%%%%%%%%%%%%%%%%%%%%%%%%%%%%%%%%%%%%%%%%%%%%%%%%%%
\subsection{Relation to Fermionized Computation}
%%%%%%%%%%%%%%%%%%%%%%%%%%%%%%%%%%%%%%%%%%%%%%%%%%%%%%%%%%%%%%%%%%%%%%%%%%%%%%%%%%

The $U (p)\times U (q)$ structure that allowed us to easily carry out the integrals in \er{rhinoceros} and \er{moose} is obscured when the fermions are not bosonized.  It is interesting to perform the same calculation without bosonizing the fermions, and then compare the two representations of the calculation.  

In the R-R sector, the fermionic correlators are equivalent to the expectation value of $2k+1 $ (holomorphic) fermions on a sphere with $\Theta_s$ at 0 and $\Theta_{s'}$ at infinity -- that is, the expectation values of $2k+1 $ holomorphic Ramond-sector fermions on a sphere.  The contractions between two timelike fermions are therefore given by
\[
D ^ {(R)} (t_1, t_2) =-\onehalf \cot\left( \frac{t_1 - t_2}{2}\right),
\]
where we have included the phase factors necessary to transform the propagator to the angular variables $t_i$.
Because we have an odd number of fermions, all terms in the correlation function will have one uncontracted fermion.  The expectation value of this last fermion is
\[
\langle \psi (t)\rangle =\sqrt{\frac{i}{2}}\,\, (2 s_0)\delta_{s_0, s'_0}\delta_{\vec{s}, -\vec{s}'},
\]
where again we have included the phase factor coming from the conformal transformation of $\psi$.

The contribution from $s_0$ again imposes that either $s_0 = s'_0 =\onehalf $ or $s_0 = s'_0 = -\onehalf $.  The factor of $2 s_0$ provides the relative sign between the two choices; this can be thought of as coming from the fact that $\psi ^ 0_0 = (\chi ^ +-\chi ^ -)/\sqrt{2}$, where $\chi$ are the raising and lowering operators for the $s$ basis. To simplify our notation, we will drop the spin structure contributions, $(2 s_0)\delta_{s_0, s'_0}\delta_{\vec{s}, -\vec{s}'}$, to the amplitude.
We find that $\mca_{2k+1} $ is given in fermionic language by
\beq
\label{eq:fermion-n}
\mca_{2k+1} =
\frac{1}{ (2k+1) !} \int_0 ^{2\pi}\prod_{i = 1} ^{2k+1}\frac{dt_i}{2\pi}
\prod_{i < j} 2\sin \left(\frac{t_i-t_j}{2}\right) 
\sum_{\sigma\in Q}\left[\sign{\sigma}
\prod_{l = 1} ^{k}\cot \left(\frac{t_{\sigma (2l-1)}-t_{\sigma (2l)} }{2}\right)\right],
\eeq
where we have again made an overall choice of phase to simplify the notation.  Here $Q$ denotes the set of all distinct contractions.

In the NS-NS sector, the contractions between the fermions appearing in the boundary perturbation are
\[
D ^{(NS)} (t_1,t_2) =-{1\over 2}\,\sin ^{-1}\left({t_1-t_2\over 2}\right),
\]
and \er{fish} becomes
\beq
\label{eq:elephant}
\mca_{2k} =
{2 ^ k\over (2k)!}
\int\prod_{i = 1} ^{2k}\frac{dt_i}{2\pi} 
\prod_{i < j} 2\sin \left( \frac{t_i-t_j}{2} \right)
\sum_{\sigma\in Q}\left[\prod_{l = 1} ^{k}
2\sin \left( \frac{t_{\sigma (2l-1)}-t_{\sigma (2l)} }{2} \right) \right]^ {-1},
\eeq 
where again $Q$ is the set of all distinct contractions.
The equivalence between the two representations of the $\psi$ fields guarantees that the integrands of \er{fermion-n} and \er{rhinoceros} are equal, as are the integrands of \er{elephant} and \er{moose}.  By rewriting these equalities in terms of $x_i = e ^{it_i}$, we arrive at the following identities between two symmetric polynomials.  The NS-NS correlation functions lead to the identity
\barray
\nonumber
\lefteqn{
2 ^ k\prod_{i < j} ^{2k} (x_i-x_j)\sum_{\sigma\in Q}\left[\sign{\sigma}
\prod_{l = 1} ^{k} (x_{\sigma (2l-1)}-x_{\sigma (2l)}) ^{-1}\right]
=}\\
\label{eq:ns-identity}
& &\sum_{\rho\in P} \left[
\prod_{\rho (i) <\rho (j)\leq k} (x_{\rho (i)}-x_{\rho (j)}) ^ 2
\prod_{k < \rho (i) <\rho ( j)\leq 2k} (x_{\rho (i)}-x_{\rho (j)}) ^ 2\right].
\earray

The R-R correlation functions lead to the identity
\barray
\label{eq:r-identity}
\lefteqn{
\prod_{i < j} ^{2k+1}(x_i-x_j)\sum_{\sigma\in Q}\left[\sign{\sigma}
\prod_{l = 1} ^{k} {x_{\sigma (2l-1)} +x_{\sigma (2l)}\over x_{\sigma (2l-1)}-x_{\sigma (2l)} }\right]
=}\\
\nonumber
& &\sum_{\rho\in P} \left[
\prod_{\rho (i) <\rho (j)\leq k+1} (x_{\rho (i)}-x_{\rho (j)}) ^ 2
\prod_{k+1 < \rho ( i) <\rho (j)\leq 2k+1} (x_{\rho (i)}-x_{\rho (j)}) ^ 2
\prod_{k+1 <\rho ( i)\leq 2k+1} x_{\rho (i)}\right].
\earray
In these forms, the identities hold for any abstract set of variables $\{x_i\} $.
We may check these results by multiplying out the terms at finite order.
For example, when $k =2$ \er{ns-identity} becomes
\barrayn
\lefteqn{
4 \left[(x_1-x_3) (x_1-x_4) (x_2 - x_3) (x_2 - x_4) - (x_1 - x_2) (x_1 -
   x_4) (x_2 - x_3) (x_3 - x_4) + \right. 
}\\
\lefteqn{
\left.  \phantom{space base} (x_1 - x_2) (x_1 - x_3) (x_2 - x_4) (x_3 - x_4)\right]
}\\
& = & 2\left[(x_1 - x_3)^2(x_2 - x_4)^2 + (x_1 - x_4)^2(x_2 - x_3)^2 + 
  (x_2 - x_1) ^ 2 (x_3 - x_4) ^ 2\right].
\earrayn
This can be seen to be true by multiplying out the terms.  Likewise, when $k = 1 $, \er{r-identity} becomes
\barrayn
\lefteqn{(x_1+x_2)(x_1-x_3) (x_2-x_3) - (x_1-x_2)(x_1+ x_3) (x_2-x_3) +(x_1-x_2)(x_1-x_3) (x_2+x_3)}\\
& \phantom{space baseasdasdfadf} &=(x_1-x_2) ^ 2x_3+(x_1-x_3) ^ 2x_2+(x_2-x_3) ^ 2x_1. \phantom{space baseasdfasfasdfasdf}
\earrayn

%%%%%%%%%%%%%%%%%%%%%%%%%%%%%%%%%%%%%%%%%%%%%%%%%%%%%%%%%%%%%%%%%%%%%%%%%%%%%%%%%%%%%
\section{General Closed String One-Point Function As a Matrix Integral}
%%%%%%%%%%%%%%%%%%%%%%%%%%%%%%%%%%%%%%%%%%%%%%%%%%%%%%%%%%%%%%%%%%%%%%%%%%%%%%%%%%%%%

In this section, we extend the results of the previous section to sketch how more general one-point functions in the rolling tachyon background can be evaluated using matrix integrals.  This is a generalization of the work that was done for the bosonic rolling tachyon background in \cite{Neil}.

Since the group structure is most evident using the bosonized representation of the $\psi$ fields, we will work in bosonized language.  A general closed string operator built out of the $\psi ^ 0$ field will take the form
\[
\mcv_f =\prod_{i} \left({1\over (\nu_i-1)!}\partial ^{\nu_i} H \right) ^{N_i}
\prod_j \left({1\over (\tnu_j-1)!}\bar{\partial} ^{\tnu_j}\tH\right)^{\tN_j}\, e ^{i (pH+ p'\tH)},
\]
where $p, p'\in \openz$ in the NS-NS sector, and $p, p'\in\openz+\onehalf$ in the R-R sector.
A general closed string operator built out of the $X ^ 0$ field will take the form
\[
\mcv_b =\prod_{i} \left({\sqrt{ 2 }\over (\mu_i-1)!}\partial ^{\mu_i} X ^ 0\right) ^{M_i}
\prod_j \left({\sqrt{ 2 }\over (\tmu_j-1)!}\bar{\partial} ^{\tmu_j} X ^ 0\right)^{\tM_j}.
\]
We have chosen here a convenient normalization for the operators.

We want to evaluate the amplitude
\beq
\label{eq:amplitude}
\left\langle \exp\left( \sqrt{ 2 }\pi\lambda\int {dt\over 2\pi}\, \eta\psi ^ 0 e ^{X ^ 0}\right)
\mcv_f\mcv_b (0)\right\rangle_{disk} =\sum_n (i\pi\tl) ^ n\mca_n.
\eeq
We note immediately that due to the boundary fermion $\eta$,  $n$ must be even in the NS-NS sector, and odd in the R-R sector.
Consider first the bosonic portion of this correlator.
The propagator for a timelike boson on the disk is
\[
G_b (z, w) =\onehalf\ln | z-w | ^ 2+\onehalf\ln | z\bar{w}-1 | ^ 2.
\]
The bosonic correlators in \er{amplitude} are then
\[
\prod_{i < j} 2\sin\left({t_i-t_j\over 2}\right)\,
\prod_k\left({1\over (\nu_k-1)!}\sum_{l = 1} ^ n\partial ^{\nu_k} G_b (0, e ^{it_l})\right) ^{N_k} \,
\prod_j\left({1\over (\tnu_j-1)!}\sum_{l = 1} ^ n\partial ^{\tnu_j} G_b (0, e ^{it_l})\right) ^{\tN_j},
\]
or
\beq
\label{eq:b-general}
\prod_{i < j} 2\sin\left({t_i-t_j\over 2}\right)\,
\prod_k\left(\sum_{l = 1} ^ ne ^{-it_l\nu_k}\right) ^{N_k} \,
\prod_j\left(\sum_{l = 1} ^ n e ^{it_l\tnu_j}\right) ^{\tN_j}.
\eeq
Meanwhile, the propagator for the $H$ field on the disk is
\[
G_f (z, w) =\ln (z-w) ^ 2
\] 
and the propagator between a $H$ field and a $\tH$ field is
\[
G_{\tilde{f}} (z, w) =\ln (z\bar{w}-1) ^ 2.
\]
The fermionic correlators in \er{amplitude} are thus
\barrayn
\lefteqn{
2 ^{-n/2}\prod_{i < j} (e ^{it_i}-e ^{it_j}) ^{\epsilon_i\epsilon_j}
\prod_{i}e ^{it_i (2p\epsilon_i +1)/2} \,\times}\\
&\times &\prod_j\left({1\over (\nu_j-1)!} \sum_l\epsilon_l\partial ^{\nu_j} G_f (0, w_l)\right) ^{N_j}
\prod_k\left({1\over (\tnu_k-1)!} \sum_l\epsilon_l\bar{\partial} ^{\tnu_k} G_{\tilde{f}} (0, w_l)\right) ^{\tN_k},
\earrayn
or
\[
\left({1\over\sqrt{ 2 }}\right) ^ n
\prod_{i < j} (e ^{it_i}-e ^{it_j}) ^{\epsilon_i\epsilon_j}
\prod_{i}e ^{it_i (2p\epsilon_i +1)/2} \,
\prod_j\left( \sum_l\epsilon_l e ^{-it_l\nu_j}\right) ^{N_j}
\prod_{k}\left(\sum_l\epsilon_le ^{it_l (\tnu_k-1)}\right) ^{\tN_j}.
\]
Here, as in the previous section, $\epsilon_i =\pm 1$, and the conservation of $H$ momentum leads to the condition
\[
\sum_i\epsilon_i +p+p' = 0.
\]
When this condition cannot be fulfilled, that is, if $| p+p' | > n$, the correlator vanishes.
When the correlator does not vanish, we will need to sum over all possible groupings of the $\epsilon_i$ into a group of $k$ with $\epsilon_i =-1$ and a group of $n-k$ with $\epsilon_i = 1$, where
\[
k =\onehalf (n +p+p').
\]  
There are $\left(\begin{array}{c} n\\k\end{array}\right)$ such terms, all of which contribute equally after integration.  Let us choose the ordering
\[
\epsilon_i =-1,\phantom{mm} 1\leq i\leq k
\phantom{space or space or}
\epsilon_i = 1,\phantom{mm} k < i\leq n.
\]
We find that the fermionic correlators become
\beq
\label{eq:f-general}
{2 ^{-n/2} n!\over k!  (n-k)!}\prod_{i < j} ^ n\left(2\sin\left({t_i-t_j\over 2}\right)\right) ^{\epsilon_i\epsilon_j}
\prod_i ^ n e ^{it_i \epsilon_i (p-p')}
\prod_j\left(\sum_l\epsilon_le ^{-it_l\nu_j}\right) ^{N_j}
\prod_{k}\left(\sum_l\epsilon_le ^{it_l (\tnu_k-1)}\right) ^{\tN_k}.
\eeq
We are now ready to put together equations \er{b-general} and \er{f-general} and write down the amplitude for a general closed string vertex operator and $n$ insertions of the boundary action.
Defining
\[
U =\diagonal{e ^{it_1},\ldots, e ^{it_k}}
\]
and
\[
V =\diagonal{e ^{it_{k+1}},\ldots, e ^{it_n}},
\]
we find
\barray
\nonumber
\mca_n & = & \int_{U (k)} [dU]\,\,\int_{U (n-k)} [dV]\,\,
\left({\det U\over\det V}\right) ^{p'-p} \times\\
\nonumber
& & \phantom{mm}
\prod_{i}\left(\Tr (U ^{\dag}) ^{\mu_i} +\Tr (V ^{\dag}) ^{\mu_i}\right) ^{M_i}
\prod_{j}\left(\Tr U ^{\tmu_j}-\Tr V ^{\tmu_j}\right) ^{\tM_j} \times\\
& &\phantom{mm}
\prod_{k}\left(\Tr (U ^{\dag}) ^{\nu_k}-\Tr (V ^{\dag}) ^{\nu_k}\right) ^{N_k}
\prod_{l}\left(\Tr U ^{\tnu_l-1}-\Tr V ^{\tnu_l-1}\right) ^{\tN_l}.
\earray
This is indeed a matrix integral with product structure $U (k)\times U (n-k)$.  In principle, the integrals over the unitary groups may be performed, as in \cite{Neil}.  However, in practice the expansions of the products above must become prohibitive, as the exponents $M_i, N_j$ can become arbitrarily large, as can the cardinality of the index sets $\{i\},\{j\}$.

%%%%%%%%%%%%%%%%%%%%%%%%%%%%%%%%%%%%%%%%%%%%%%%%%%%%%%%%%%%%%%%%%%%%%%%%%%%%%%%%%%%%%
\section{Closed String Production}
%%%%%%%%%%%%%%%%%%%%%%%%%%%%%%%%%%%%%%%%%%%%%%%%%%%%%%%%%%%%%%%%%%%%%%%%%%%%%%%%%%%%%

In this section, we calculate closed string emission from the decaying D-branes described by the boundary state we derived in section 2.  Analogous computations were done for the bosonic string in \cite{llm, klms}.  The computation for the superstring does not differ greatly from that done for the bosonic string, and our discussion will therefore be largely parallel to that in \cite{llm}.

We begin with the expression for the cylinder diagram
\beq
\label{eq:w}
W ={1\over 2}\left\langle B \left| {1\over L_0+\widetilde{L}_0+ i\epsilon}\right| B\right\rangle.
\eeq
This expression picks up an imaginary part whenever particles go on shell.  Writing 
\[
\ket{B} =\sum_sU (\omega_s)\ket{s},
\]
the imaginary part of \er{w} is
\beq
\label{eq:imw}
\mbox{Im}\, W ={1\over 2}\sum_s{1\over 2\omega_s} | U (\omega_s) | ^ 2.
\eeq
The optical theorem relates this to particle production, so that $\bar{N}/V = 2\,\mbox{Im}\, W$.  The volume $V$ here is the $p$-dimensional spatial volume of the decaying brane.

We first need to calculate $U (\omega)$.  This reduces essentially to choosing a prescription for integrating the zero mode of the bosonic coordinate. Assuming the factorization \er{vns} and \er{vr} for a general closed string vertex operator, the computation of $U (\omega)$ factorizes into one part from the expectation value of $\mcv ^ {\perp} (X ^ i,\psi ^ i,\psit ^ i)$ in the transverse CFT, and the part of interest from the time-dependent components of the boundary state.  As in \cite{llm}, the transverse CFT contributes only a phase, which drops out of the computations below.

The contribution from the $(X ^ 0,\psi ^ 0) $ CFT takes the form
\beq
\label{eq:integral-def}
U (E) =\langle e ^{i EX ^ 0}\rangle_{disk} = i\int_{\mcc} dt\, g (t) e ^{i Et}.
\eeq
Here $g (t) $ is the function appearing in the lowest component of the boundary state, which we computed for the rolling tachyon profile in section 2. This function is given by \er{gns-def} in the NS-NS sector, and \er{gr-def} in the R-R sector.  For the bouncing tachyon profile, $g (t)$ is given by \er{fns} in the NS-NS sector, and \er{fdef} in the R-R sector.   

There are two contours of interest, namely $\mcc_{real}$, which runs along the real axis and is closed in the upper half plane, and the Hartle-Hawking contour $\mcc_{H H}$, which runs along the imaginary axis from $t = i\infinity$ to $t = 0$, then runs along the real axis \cite{llm}. For the rolling tachyon profile, both contours yield the same result.  For the bouncing tachyon profile, the contour $\mcc_{H H}$ yields the same result as for the rolling tachyon, while the contour $\mcc_{real}$ yields a different result.

We will consider the rolling tachyon profile first. For this profile, the contour integral \er{integral-def} yields in the NS-NS sector
\[
U_{NS} (\omega) =
i\int_{\mcc_{real}} dt\, {1\over 1+\pi ^ 2\lambda ^ 2e ^{\sqrt{2} t}}\, e ^{i\omega t} = 
(\pi\lambda) ^{-i\sqrt{ 2 }\omega}\,{\pi/\sqrt{ 2 }\over\sinh {\pi\omega\over\sqrt{ 2 }}}.
\]
In the Ramond sector, the contour integral calculation yields
\[
U_{R} (\omega) =
i\int_{\mcc_{real}} dt\, {i\pi\lambda e ^{t/\sqrt{ 2 }}\over 1+\pi ^ 2\lambda ^  2e ^{\sqrt{2} t}} e ^{i\omega t} =
- (\pi\lambda) ^{-i\sqrt{ 2 }\omega}\,{\pi/\sqrt{ 2 }\over\cosh {\pi\omega\over\sqrt{ 2 }}} .
\]
We now use these results for $U (\omega)$ to calculate closed string emission from the decaying branes, using \er{imw}.

Let us consider first the portion of the cylinder diagram coming from the overlap $\left._{NS}\bra{B;+}{ 1\over L_0+\widetilde{L}_0+ i\epsilon}\ket{B; +}_{NS}\right.  $.  We have argued that the imaginary part of this amplitude is given by 
\[
2\,\mbox{Im } W_{NS++} =\sum_s{1\over 2\omega_s} | U (\omega_s) | ^ 2 =\sum_s{\pi ^ 2\over 4\omega_s}{1\over\sinh ^ 2  (\pi\omega_s/\sqrt{ 2 })}.
\]
We now go through a series of manipulations to bring this into a more familiar form. Using
\[
{1\over\sinh ^ 2 (\pi\omega_s/\sqrt{ 2 })} = 4\sum_{n = 0} ^{\infinity} n\, e ^{-\sqrt{ 2 }\pi n\omega_s}
\]
and
\barray
\nonumber
{1\over\sqrt{ 2 }\omega_s} e ^{-\sqrt{ 2 }\pi\omega_s n} & = &
	{1\over 2\pi}\int dk_0\,{1\over k_0 ^ 2+ {1\over 2}\omega_s ^ 2}\, e ^{2\pi ik_0n}\\
\label{eq:zebra}
	& = &{1\over 2\pi}\int dk_0\int dt\, e ^{-t (k_0 ^ 2+{1\over 2}\omega_s ^ 2)} e ^{2\pi ik_0n}\\
\nonumber
	& = &{1\over 2\pi}\int dt\,\sqrt{{\pi\over t}} \, e ^{-\pi ^ 2n ^ 2/t} e ^{-t\omega_s ^ 2/2},
\earray
the contribution to particle production is
\[
2\,\mbox{Im } W_{NS++} = C_p\sum_{n} n\int dt\, t ^{-1/2} e ^{-\pi ^ 2n ^ 2/t} \sum_s e ^{-t\omega_s ^ 2/2}.
\]
Here $C_p$ is a numerical constant coming from the normalization of the boundary state. We now write $\omega_s ^ 2 = 4N_{\psi} + 4N_X-2$ and set $q= e ^{-2t}$.  The sum over states can now be performed in the usual manner, so that
\[
2\,\mbox{Im } W_{NS++} = C_p\sum_nn\int dt\, t ^{-1/2} e ^{-\pi ^ 2n ^ 2/t}\left({f_3 (q )\over\eta (q)}\right) ^ 8.
\]
In a very similar fashion, one may calculate the contribution to particle production from the overlaps $\left._{NS}\bra{B;+}{ 1\over L_0+\widetilde{L}_0+ i\epsilon}\ket{B; -}_{NS}\right. $ and $\left._R\bra{B;+}{ 1\over L_0+\widetilde{L}_0+ i\epsilon}\ket{B; +}_R\right.  $.  The total number of emitted particles is then
\[
{\bar{N}\over V} =
C_p\sum_nn\int dt\, t ^{-1/2} e ^{-\pi ^ 2n ^ 2/t} [\eta (q)] ^{-8}\left( f_3 ^ 8 (q )-f_4 ^ 8 (q)-(-1) ^ nf_2 ^ 8 (q)\right).
\]
This expression diverges for small $t$.  This divergence is easier to examine after making a modular transformation to $s ={\pi\over t}$. Defining $w = e ^{-2\pi s}$, we find
\[
{\bar{N}\over V} =\sum_n n\int {ds\over s}\int d ^ 9 k\, w ^{-(k ^ 2+n ^ 2)/2} [\eta (w)] ^{-8}\left(f_3 ^ 8 (w)-(-1) ^ n f_4 ^ 8 (w)-f_2 ^ 8 (w)  \right).
\]
For a decaying D$p$-brane, the answer generalizes in the obvious fashion,
\beq
\label{eq:imw-result}
{\bar{N}\over V} =\sum_n n \int {ds\over s}\int d ^ p k\, w ^{-(k ^ 2+n ^ 2)/2} [\eta (w)] ^{-8}\left(f_3 ^ 8 (w)-(-1) ^ n f_4 ^ 8 (w)-f_2 ^ 8 (w)  \right)
\eeq
When $s\to\infinity$, we may expand \er{imw-result} for small $w$ to find
\[
\int {ds\over s}\,\int d ^ pk\; w ^{k ^ 2/2}\, (1+\mco (w ^{-1/2}))\sim\int{ds\over s} s ^{-p/2}.
\]
It is now easy to see that this diverges for $p = 0$.  
As noted in \cite{llm}, this has the following open string interpretation.  Let
\[
\ket{B_+} =\sum_{n = 0} \ket{B_p; (-1) ^ n; x = \sqrt{ 2 }(n+\onehalf)\pi}
\]
where $\ket{B_p; (-1) ^ n; x = \sqrt{ 2 }(n + \onehalf)\pi}$ denotes the boundary state for a Euclidean D-instanton with $p$ spatial dimensions, located at the position $x = \sqrt{ 2 }(n+\onehalf)\pi$ in the Euclidean time direction, sourcing positive or negative R-R flux according to the sign $(-1) ^ n$.  Analogously, we define
\[
\ket{B_-} =\sum_{n = 0}\ket{B_p;-(-1) ^ n; x =-\sqrt{ 2 }(n+\onehalf)\pi}.
\]
The imaginary part of the partition function \er{imw-result}, is then
\[
{\bar{N}\over V} =\left\langle B_-\left|{1\over L_0+\widetilde{L}_0} \right| B_+\right\rangle.
\]
In other words, the expression \er{imw-result} is the partition function that arises from a collection of alternating Euclidean D$p$-branes and anti-D$p$-branes located at regular positions along the imaginary time axis, provided we only include the open strings that stretch across the line $x = 0$ \cite{llm}.  
This open string description of the closed string radiation arises from writing the closed string fields produced by the decaying D-brane as radiation sourced by D-instantons in Euclidean time.  From this point of view, the Euclidean path integral over $x < 0$ in the presence of the source $\ket{B_-}$ prepares a state in the closed string Hilbert space at $x = 0$ which is closely related to the final closed string state; see \cite{llm} for a complete discussion.

In this open string language, the divergence in equation \er{imw-result} comes from the existence of a massless string state stretching between the instantons nearest to the imaginary time axis.  This divergence comes from the IR of the open string channel, and thus from the UV of the closed string channel.  It reflects the fact that the amplitude for emission of a very massive string state does not fall off quickly enough as a function of energy to offset the Hagedorn divergence in the number of states.  While this divergence only occurs for $p = 0$, the homogenous solution studied here should not be trusted for larger $p$, as higher dimensional D-branes should rapidly break up into causally disconnected patches that decay locally as unstable D0-branes.  (This discussion is complicated slightly in the superstring by the R-R couplings, but it is easy to see that the NS-NS contributions to particle production are divergent on their own for $p=0$.)  Therefore the divergence that occurs when $p = 0$ should be thought of as the generic tree level result in the superstring as well as the bosonic string.  This seems to signal the breakdown of string perturbation theory.

We may also consider the energy that is radiated into these closed string modes,
\[
{E\over V} =\sum_s{1\over 2} | U (\omega_s) | ^ 2.
\]
Since from \er{zebra} we may write
\[
-{1\over 2\pi}\,\dda\left[{1\over\sqrt{ 2 }\omega_s}\, e ^{-\sqrt{ 2 }\pi\omega_s (n+a)}\right]_{a = 0} =
{1\over 2} e ^{-\sqrt{ 2 }\pi\omega_sn} ={1\over 2\pi}\int dt\left({\pi\over t}\right) ^{3/2} ne ^{-t\omega_s ^ 2/2},
\]
the energy radiated into closed strings is given by
\beq
\label{eq:energy-radiated}
{E\over V} =\sum_nn ^ 2\int{ds\over s}\,\int d ^ pk\, s\, w ^{(k ^ 2+n ^ 2)/2} [\eta (w)] ^{-8}\left(f_3 ^ 8 (w)-(-1) ^ n f_4 ^ 8 (w)-f_2 ^ 8 (w)  \right).
\eeq
The divergence at small $w$ has now been worsened,
\[
{E\over V}\sim\int ds\, s ^{-p/2},
\]
so that the total energy emitted into closed strings is divergent for $p\leq 2$.
This is precisely the expected supersymmetric generalization of the results of \cite{llm} for the bosonic string, and suggests that all the energy of the D-brane is converted into closed strings.

The story is slightly more complicated for the bouncing tachyon profile when the contour $\mcc_{real}$ is used in equation \er{integral-def}, but the fundamental physical picture does not change.  With this contour, we find
\barrayn
U (\omega)_{NS} & = & {-\sqrt{ 2 }\pi\over\sinh (\pi\omega/\sqrt{ 2 })}\,\sin (\sqrt{ 2 }\omega\kappa),\\
U (\omega)_R & = &{\sqrt{ 2 }\pi\over\cosh (\pi\omega/\sqrt{ 2 })}\, \cos (\sqrt{ 2 }\omega\kappa),
\earrayn
where $\kappa\equiv\ln\sin\pi\lambda$.  
The number of emitted particles is now
\barrayn
{\bar{N}\over V} &=&
     \sum_{n = 0}  n\int{ds\over s}\int d ^ pk\,w ^{k ^ 2/2} [\eta (w)] ^{-8} \left( 2w ^{n ^ 2/2}\left(f_3 ^ 8 (w)-(-1) ^ n f_4 ^ 8 (w)-f_2 ^ 8 (w)  \right)\right. -\\
& & \phantom{a skeptical}
\left. \left( w ^{-(n+2i\kappa /\pi) ^ 2/2} +w ^{-(n+2i\kappa /\pi) ^ 2/2} \right)
\left(f_3 ^ 8 (w) + (-1) ^ n f_4 ^ 8 (w)-f_2 ^ 8 (w)  \right) \right).
\earrayn
Note that the signs of the GSO projection are reversed in the second line of the above equation.  Due to this change of sign, the second line is finite.  However, the first line is identical (up to an overall factor of 2) to the calculation performed for the rolling tachyon profile, and in particular diverges in the same way.

%%%%%%%%%%%%%%%%%%%%%%%%%%%%%%%%%%%%%%%%%%%%%%%%%%%%%%%%%%%%%%%%%%%%%%%%%%%%%%%%%%%%%
\section{Conclusions}
%%%%%%%%%%%%%%%%%%%%%%%%%%%%%%%%%%%%%%%%%%%%%%%%%%%%%%%%%%%%%%%%%%%%%%%%%%%%%%%%%%%%%

We have presented a basic analysis of closed string one-point functions in rolling tachyon backgrounds in Type II string theories.  We noted the appearance of an interesting $U (p)\times U (q)$ structure that allowed us to easily evaluate the integrals appearing in the correlation functions, and pointed out how this structure could be generalized to compute one-point functions involving excitations of the $X ^ 0$ and $\psi ^ 0$ oscillators.  We confirmed that the total number and energy of emitted particles diverge for low-dimensional D-branes.

The computations of closed string expectation values in the rolling tachyon background could be extended to multi-point amplitudes of several closed string operators on the disk. These amplitudes should also be computable in terms of matrix integrals. It would be interesting to understand if there is any deeper meaning to the matrix integral structure appearing in the closed string correlators in the rolling tachyon background.

One of the biggest outstanding problems in this approach to D-brane decay is the proper way to handle the divergences that appear in the calculation of the number and energy of the closed strings radiated from the brane.  Naively, it seems as if the open string physics must be modified at late times.   The divergence in equations \er{imw-result} and \er{energy-radiated} comes from the IR of the open string channel, suggesting that the proper way to address the divergence is to shift the open string background.  The parts of the rolling tachyon boundary state that contain $X ^ 0,\psi ^ 0$ oscillators all grow exponentially at late times \cite{grow}, a divergence which does not appear at tree level, but affects physical quantities through open string loop diagrams.  This suggests that the late time behavior of the open string solution as represented in the boundary state must be modified.  The correct way to modify the open string solution remains unclear; perhaps the matrix model may offer some guidance.

Also, since we have a clear picture of the coherent closed string state that the brane decays into, it is natural to ask whether there is an open string description of this state.  This is certainly true in two-dimensional string theory \cite{kms}, and it would be of great interest to work out such a description in the full 10-dimensional theory. Once this open string description is found, the necessary steps to remove the divergence may well be more evident.  These are all issues deserving of further study.

\medskip
\medskip

\noindent {\large {\bf Acknowledgments}}

\noindent The author wishes to thank I.~Ellwood, S.~Robinson, W.~Taylor, and especially H.~Liu for useful conversations.  Thanks to H.~Liu for suggesting this topic, and to H.~Liu and W.~Taylor for comments on the manuscript.  Research supported in part
by the U.S. Department of Energy 
under cooperative research
agreement $\#$DF-FC02-94ER40818.

\appendix

%%%%%%%%%%%%%%%%%%%%%%%%%%%%%%%%%%%%%%%%%%%%%%%%%%%%%%%%%%%%%%%%%%%%%%%%%%%%%%%%%%%%%
\section{An Alternate Derivation of the Boundary State: the SU(2) Method}
%%%%%%%%%%%%%%%%%%%%%%%%%%%%%%%%%%%%%%%%%%%%%%%%%%%%%%%%%%%%%%%%%%%%%%%%%%%%%%%%%%%%%

In this section we will extend Sen's calculations of \cite{sen-tm} to obtain the lowest components of the boundary states for both the rolling and bouncing tachyon backgrounds in the Ramond Ramond sector by computing the boundary state in the Euclidean theory first, and then Wick rotating.  This approach to tachyon condensation was introduced by Sen in \cite{sen-rt}.  Free conformal field theories deformed by a boundary perturbation taken from a local SU(2) current algebra were studied in \cite{cklm, rs}.  The results appearing in this section are not necessarily new (the result \er{fdef} appeared as a guess in \cite{sen-guess}), but there is no derivation of these results in the literature.

In our conventions, the SU(2) current algebra is given by
\barrayn
J^1 (z) & = & -\sqrt{2}\psi\sin (\sqrt{2} X_L (z))\\
J^2 (z)& = &\sqrt{2}\psi\cos (\sqrt{2} X_L(z))\\
J^3(z) & = & i\sqrt{2}\partial X(z).
\earrayn
We will suppress Lorentz indices whenever possible.  Here only the holomorphic part of the field $X$ is used to define the SU(2) generators.

We are interested in a set of closed string disk one-point functions in the tachyon backgrounds \er{rolling-tachyon} and \er{bouncing-tachyon}.  The key step in the argument we will use to evaluate these one-point functions is the reinterpretation of a tachyon insertion integrated around the boundary of the disk as a holomorphic contour integral of one of the elements of the SU(2) algebra above.  When the contour surrounds a vertex operator inserted in the bulk, performing the contour integral reduces to rotating the vertex operator in a sense we will make precise below.  Concretely, if the perturbation is given by $\Delta S =-\theta\int \frac{ dt}{2\pi}\, J$, the one-point function of the operator $\mc{O}(z) \tO (\bar{z})$ on the disk is
\[
\langle\mc{O}(z) \tO (\bar{z}) \rangle_{\mathrm{perturbed}} =\langle e ^{\theta\int \frac{ dt\,}{2\pi} J } \mc{O} (z) \tO (\bar{z}) \rangle_{\mathrm{unperturbed}} .
\]
We may now imagine deforming the $J$ integration contours away from the boundary one by one.  If the OPE of $\mc{O}$ and $J$ is well-defined, then $\mc{O}$ will have a well-defined transformation property with respect to the symmetry algebra to which $J$ belongs.  Thus we may conclude
\[
\langle\mc{O}\tO \rangle_{\mathrm{perturbed}} =\langle \mc{R}_{\theta} ( \mc{O})\tO \rangle_{\mathrm{unperturbed}},
\]
where $\ket{\mc{R}_{\theta} (\mc{O})} = e ^{i\theta J_0}\ket{\mc{O}}$.  For a complete discussion of these ideas, we refer the reader to \cite{rs}.  There is some subtlety involved in deforming the integration contours of boundary perturbations off the boundary.  However, in our case there is no ambiguity involved, as all operators $\mc{O}$ that we need to consider are well-defined with respect to the boundary perturbation.  In order to carry out this program,  our perturbation must be self local; in particular, it must commute with itself.  It is simple to check that, with $\eta$ included in the vertex operator, our deformations satisfy this criterion. 

In the NS-NS sector, the operators which have nonvanishing one-point functions in the rolling tachyon background are of the form $\mc{V}(z,\bar{z}) =\mc{O}_n (z) \tO_n (\bar{z}) \equiv  e ^{in\sqrt{2} X_L} e^{in\sqrt{2} X_R} $ \cite{sen-tm}.  The holomorphic part $\mc{O}_n (z)= e ^{in\sqrt{2} X_L}$ of such vertex operators transforms in the $j =| n|, m = n$ representation of SU(2).  

In the our our sector, we will be interested in operators of the form $\mc{V} =\mc{O}_{n,s} (z) \tO_{n, s'} (\bar{z}) \equiv  \Theta_s e^ {ik_n X_L}\widetilde{\Theta}_{s'} e^{ik_nX_R}$, where 
\[
\label{eq:k_n}
k_n\equiv (n +\onehalf)\sqrt{2}.
\]
The operators $\mc{O}_{n, s}$ transform in spinor representations of SU(2), $j =|n+\onehalf |, m = n+ \onehalf $.  There is a branch cut in the OPE of $J^1, J^2$ with $\mc{O}_n$ due to the half integral momentum, which offsets the branch cut the spin field puts in the fermion field; thus all contour integrals that appear are well-defined.

Because we need to group the fermions into pairs in order to define the $\Theta_s$, we will need to carry around a ``spectator dimension''; in other words, while we are perturbing only one spatial direction $\mu$, we must always consider both the $\mu$ and the $\mu + 1$ directions in order to be able to write down the spin fields.  We will see that in the end, however, the dependence of the boundary state on the fermion zeromodes will indeed factorize.

\subsection{Bouncing Tachyon Profile}

We first consider the tachyon profile $T (X) =\lambda\cos (\sqrt{2} X)$.  For this tachyon profile, the boundary action is proportional to $J^1$:
\[ 
\Delta S = -\pi\lambda\int \,\frac{ dt}{2\pi} J^1.  
\]
Using the state-operator correspondence, we may write
\[
e ^ {\pi\lambda\int \,\frac{ dt}{2\pi} J^1} \mc{O}_{n, s}
      \cong e ^{i\pi\lambda J^1_0}\ket{\mc{O}_{n, s}},
\]
so the boundary perturbation becomes a rotation of $\ket{\mc{O}}$ by $\pi\lambda$ around the 1 axis; we denote the group element of SU(2) corresponding to this rotation as $\Gamma$.  Now $J^1\propto J_++J_-$, so action of $J^1$ will, in general, mix $\mc{O}_{n, s}$ with all the other operators in its SU(2) multiplet.  However, once we take the expectation value, only operators whose bosonic part is of the form $e ^{ik_n (X_R- X_L)}$ will contribute.  This convenient fact is due to conservation of momentum on the disk with Neumann boundary conditions \cite{sen-tm}.  For our purposes, we therefore only need to know that a rotation by $J^1$ through an angle of $\pi\lambda$ will take the highest (lowest)-weight state $\Theta_s e ^{ik_n X_L}$ to the lowest (highest)-weight state $\Theta_{-s} e ^{-ik_n X_L}$ multiplied by some calculable coefficient plus operators whose expectation values vanish.  The coefficient we need is given by $D^{j=|n+1/2|}_{n+1/2,-n-1/2} (\Gamma) = (i\sin (\lambda\pi))  ^{2| n +\onehalf |}$, \cite{rs}.

The lowest component of the boundary state $\ket{B, +}$ will take the form
\[
\ket{B, +}_R =\sum_n f_n\ket{k_n; -+} +\mathrm{oscillators},
\]
where the $f_n$ are c-numbers. The periodicity of the potential ensures that the restriction to momenta of the form $k_n$ is legitimate.  The choice of spinor vacuum $\ket{-+}$ is what we expect for a decaying unstable D9-brane, as it has the right Lorentz properties to source $C_8$ flux.  We will demonstrate below that $\ket{-+}$ is indeed the correct choice.

To solve for $f_n$, we have
\[
f_n =\langle k_n; -+\ket{B, +} =\langle\mc{O}_{-+} e ^{ik_nX}\rangle_{perturbed} =
\langle\mc{R}_{\Gamma} (\mc{O}_{-+}e ^{ik_nX})\rangle_{unperturbed}.
\]
Here $\Theta_{-+}$ is the combination of spin field operators corresponding to the state $\ket{-+}$.  We can use our spinor dictionary \er{spin-spatial} to find
\barrayn
\mc{R}_{\Gamma} (\Theta_{-+} e^{ik_n X_L}) & = & D^{j=|n+1/2|}_{n+1/2,-n-1/2} (\Gamma)\Theta_{++} e^{-ik_n X_L} +\ldots \\
\mc{R}_{\Gamma} (\Theta_{+-} e^{ik_n X_L}) & = &D^{j=|n+1/2|}_{n+1/2,-n-1/2} (\Gamma)\Theta_{--}e^{-ik_n X_L} +\ldots .
\earrayn
Note that the rotated operator has the right quantum numbers to have nonvanishing one point function on the disk.  Note also that the effect of the perturbation on the spinor vacuum has indeed factorized as advertised; we will drop the spectator dimension from this point on.
Therefore,
\beq
f_n  =(i\sin (\lambda\pi))^{2| n +\onehalf |}
\left\langle \Theta_{++} e ^{ik_n (X_R- X_L)}  \right\rangle_{unperturbed} .
\eeq
The expectation value $\langle \Theta_{++} e ^{ik_n (X_R- X_L)}  \rangle$ contributes only an overall constant, and the boundary state now becomes, up to normalization,
\[
\ket{B,+} =\sum_{n\phantom{v}\mathrm{odd}} \left[ (i\sin (\lambda\pi))^{2| n +\onehalf |} \,\epsilon (n) \, e ^{ik_n X} \right]\ket{0;-} +\ldots .
\]
Here $\epsilon (n)$ is a phase factor which we must include because the basis for which the coefficient $D^{j=|n+1/2|}_{n+1/2,-n-1/2}$ is calculated may not be the same basis that we are using \cite{sen-tm}.  This phase factor may be fixed by asking that when $\lambda =\onehalf$, the boundary state reduce to that of alternating D-branes and anti-D-branes placed at $a_i =\sqrt{2}\pi (n+\onehalf)$.  Using this criterion, we find $\epsilon (n) =\sign{n} $.  

We may now perform the sum on $n$:
\[
 \sum_{n} \left[ (i\sin (\lambda\pi))^{2| n +\onehalf |} \epsilon (n) e ^{ik_n X} \right]  = 
\sin (\lambda\pi)
\left[\frac{e ^ {iX/\sqrt{2}} }{1+\sin ^ 2 (\lambda\pi) e ^{ i\sqrt{2} X}} -
\frac{e ^{ -iX/\sqrt{2}} }{1+\sin ^ 2 (\lambda\pi) e ^ {-i\sqrt{2} X}} 
\right] .
\]
At this point we may Wick rotate: setting
\beq
\label{eq:fdef}
f_R (X ^ 0) = \sin (\lambda\pi)
\left[\frac{e ^ {X ^ 0/\sqrt{2}} }{1+\sin ^ 2 (\lambda\pi) e ^{ \sqrt{2} X ^ 0}} -
\frac{e ^{ -X ^ 0/\sqrt{2}} }{1+\sin ^ 2 (\lambda\pi) e ^ {-\sqrt{2} X ^ 0}} 
\right],
\eeq
the boundary state for the $X ^ 0$ direction becomes
\[
\ket{B; +}_R = f_R (X ^ 0)\ket{0; -} +\ldots,
\]
while
\[
\ket{B; -}_R =f_R (X ^ 0) \ket{0; +} +\ldots.
\]
Our result for $f_R (X^0)$ agrees with that written down in \cite{sen-guess}.

The corresponding formula for the NS-NS sector was found in \cite{sen-tm}.  Here
\[
\ket{B;\eta}_{NS} = f_{NS} (X ^ 0)\0 +\mathrm{oscillators},
\]
where
\beq
\label{eq:fns}
f_{NS} (X ^ 0) =\frac{1}{1+\sin ^ 2 (\lambda\pi) e ^{\sqrt{2}X^0}} +\frac{1}{1+\sin ^ 2 (\lambda\pi) e ^{-\sqrt{2}X^0}} -1 .
\eeq

\subsection{Rolling Tachyon Profile}

We now turn our attention to the tachyon profile $T (X) =\lambda e ^{iX/\sqrt{2}}$.  The perturbing boundary action for this profile is proportional to the raising operator of the SU(2) algebra,
\[
S_{boundary} = -\pi\lambda\int\frac{dt}{2\pi} J ^ + .
\]
We obtain the lowest components of the boundary state in both the NS-NS and the R-R sectors.  This serves as a check on our computations in section 2.

Once again, matters are simplified by the fact that only the lowest and highest weight states have nonvanishing expectation values, as a consequence of Neumann boundary conditions.  Therefore we only need to consider operators of the form $\mc{V} =\mc{O}_{j, m = - j}\tO$, and the effect of the boundary perturbation is simple:
\[
e ^{i\pi\lambda\int\frac{dt}{2\pi} J ^ +}\mc{O}_{j,m = -j} = (i\pi\lambda) ^ {2j}\mc{O}_{j,m = j}.
\]
Therefore, the lowest component of the NS-NS sector boundary state is
\[
\sum_{n =1} (i\pi\lambda) ^{2n}\ket{k_{- n} = -n\sqrt{2}}.
\]
We can perform the sum to find
\[
\ket{B,\eta }_{NS} = \frac{1}{1+\pi ^ 2\lambda ^ 2e ^{- i\sqrt{2} X}}\0 + \ldots.
\]
Upon Wick rotation we find
\[
\ket{B,\eta }_{NS} = \frac{1}{1+\pi ^ 2\lambda ^ 2e ^{\sqrt{2} X^0}}\0 + \ldots \equiv g_{NS} (X^0) \0 + \ldots.
\]
Similarly, the lowest component of the Ramond sector boundary state is
\[
\ket{B;+}_R = \sum_{n =1}(i\pi\lambda) ^{2n-1}\ket{k_{- n} = (-n+\onehalf) \sqrt{2}; -}.
\]
The dependence on the spin structure is precisely the same as that in the bouncing tachyon case above.  We can perform the sum to find
\[
\ket{B; +}_R =\frac{i\pi\lambda e ^{-iX/\sqrt{2}}}{1+\pi ^ 2\lambda ^ 2 e ^{- i \sqrt{2} X}}\ket{0;-} +\mathrm{oscillators}.
\]
Upon Wick rotation we find
\[
\ket{B; +}_R =\frac{i\pi\lambda e ^{X^0/\sqrt{2}}}{1+\pi ^ 2\lambda ^ 2 e ^{  \sqrt{2} X^0}}\ket{0;-} + \ldots \equiv g_R (X^0) \ket{0;-} +\ldots.
\]
These results agree precisely with those we found in \er{gr-def} and \er{gns-def} by directly evaluating the correlators.

%%%%%%%%%%%%%%%%%%%%%%%%%%%%%%%%%%%%%%%%%%%%%%%%%%%%%%%%%%%%%%%%%%%%%%%%%%%%%%%%%%%%%
\section{Relations between Spinor Vacua}
%%%%%%%%%%%%%%%%%%%%%%%%%%%%%%%%%%%%%%%%%%%%%%%%%%%%%%%%%%%%%%%%%%%%%%%%%%%%%%%%%%%%%

In this appendix, we summarize the relationships between the spin states $\ket{s}\ket{\tilde{s'}}$, in terms of which the Ramond Ramond spin fields have a simple description, and the spin states $\ket{\pm\pm}$, in terms of which the boundary state has a simple description.  

We define two different sets of raising operators,
\begin{eqnarray}
\chi_a ^{\dag} & = &\frac{1}{\sqrt{2}} (\psi ^{2a- 1}_0+i\psi^{2a}_0) \\
\zeta ^{2a}_+ & = &\frac{1}{\sqrt{2}}(\psi ^{2a}_0+i\psit^{2a}_0),
\end{eqnarray}
where the directions $2a -1$ and $2a$ are spatial.  Their Hermitean conjugates are similarly defined.
The $\chi_a $ are raising and lowering operators for the $\ket{s}$ basis, defined so that $\chi_a \ket{-\onehalf} = 0$, while $\chi_a ^{\dag}\ket{- \onehalf} =\ket{\onehalf}$. The $\zeta$s are raising and lowering operators for the $\ket{\pm}$ basis, $\zeta ^{2a}_-\ket{-} = 0$, $\zeta ^{2a}_+\ket{-} =\ket{+}$.

We may express the $\zeta$s in terms of the $\chi$s; for example,
\[
\zeta^{2a -1}_+ = \onehalf (\chi ^{\dag}_a +\chi_a+i (\chit ^{\dag}_a+\chit_a)) .
\]
The last piece of information we need concerns the action of the $\chit$s on the states $\ket{s}\ket{\tilde{s'}}$.  Since the operators $\chit$ need to pass through the holomorphic states $\ket{s}$ before they can act on $\ket{\tilde{s'}}$, we need to know if they pick up a sign in doing so.  In other words, we need to know when to count the state $\ket{s}$ as fermionic, which amounts to knowing the action of $(- 1) ^ F$ on the state $\ket{s}$.  On the zero modes the action of $(- 1) ^ F$ reduces to that of the chirality matrix $\Gamma$, which in the basis we are using for the ten-dimensional spinors is simply $\Gamma \ket{s_0;s_1;\ldots;s_5} = \sign{\prod s_a}\ket{s_0;s_1;\ldots;s_5}$.  Thus we assign $(- 1) ^ F\ket{-\onehalf} = - \ket{-\onehalf}$, while $(- 1) ^ F\ket{\onehalf} = \ket{\onehalf}$.

We are now ready to write down the relationship between the two bases.  We have
\begin{eqnarray}
\nonumber
\ket{++} & = &\frac{1}{\sqrt{2}} \left(\ket{\onehalf}\ket{-\onehalf} - i\ket{-\onehalf}\ket{\onehalf}\right) \\
\label{eq:spin-spatial}
\ket{--} & = &\frac{1}{\sqrt{2}} \left(\ket{-\onehalf}\ket{\onehalf} - i\ket{\onehalf}\ket{-\onehalf}\right) \\
\nonumber
\ket{+-} & = &\frac{1}{\sqrt{2}} \left(\ket{\onehalf}\ket{\onehalf} - i\ket{-\onehalf}\ket{-\onehalf}\right) \\
\nonumber
\ket{-+} & = &\frac{1}{\sqrt{2}} \left(\ket{-\onehalf}\ket{-\onehalf} -i\ket{\onehalf}\ket{\onehalf}\right) 
\end{eqnarray}
The phase of $\ket{++}$ has been determined arbitrarily, which fixes the phases of the other three states. 
When $a = 0$, the relations are altered slightly, becoming instead:
\begin{eqnarray}
\nonumber
\ket{++} ^ 0 & = &\frac{1}{\sqrt{2}} \left(\ket{\onehalf}\ket{-\onehalf} ^ 0 - i\ket{-\onehalf}\ket{\onehalf} ^ 0\right) \\
\label{eq:spin-timelike}
\ket{--} ^ 0 & = &\frac{1}{\sqrt{2}} \left(\ket{-\onehalf}\ket{\onehalf} ^ 0 - i\ket{\onehalf}\ket{-\onehalf} ^ 0\right) \\
\nonumber
\ket{+-} ^ 0 & = &\frac{1}{\sqrt{2}} \left(i\ket{\onehalf}\ket{\onehalf} ^ 0 - \ket{-\onehalf}\ket{-\onehalf} ^ 0\right) \\
\nonumber
\ket{-+} ^ 0 & = &\frac{1}{\sqrt{2}} \left(\ket{\onehalf}\ket{\onehalf} ^ 0-i\ket{-\onehalf}\ket{-\onehalf} ^ 0\right) 
\end{eqnarray}

\end{document}